\documentclass[preprint]{aastex}

\usepackage{amsmath}
\usepackage{epstopdf}

\begin{document}

\title{Triggered Star Formation and Its Consequences}
\author{Shule Li\\ Adam Frank, and Eric G. Blackman}
\affil{Department of Physics and Astronomy, University of Rochester, Rochester, NY, 14627}
\email{shuleli@pas.rochester.edu}

\begin{abstract}
Star formation can be triggered by compression from wind or supernova driven shock waves that sweep over molecular clouds. Because these  shocks will likely contain processed elements, triggered star formation has been proposed as an explanation for short lived radioactive isotopes (SLRI) in the Solar System. Previous studies have  tracked the triggering event to the earliest phases of collapse and have focused on the shock properties required for both successful star formation and mixing  of SLRI's.  In this paper, we use Adaptive Mesh Refinement (AMR) simulation methods, including sink particles, to  simulate the full collapse and  subsequent evolution of a stable Bonnor-Ebert sphere subjected to a shock and post-shock wind. We track the flow of the cloud material after a star (a sink particle) has formed.  For non-rotating clouds we find robust triggered collapse and little bound circumstellar material remaining around the post-shock collapsed core. When we add initial cloud rotation we observe the formation of disks around the collapsed core which then interact with the post-shock flow.  Our results indicate that these circumstellar disks  are massive enough to  form planets  and are  long-lived, in spite of the ablation driven by post-shock flow  ram pressure.  As a function of the initial conditions, we also track the time evolution of the accretion rates and  particle mixing between between the ambient wind and cloud material.
The latter is maximized for cases of highest mach number.
\end{abstract}

\keywords{hydrodynamics, planetary nebula, accretion disks, molecular clouds, star formation}

\section{Introduction}

Triggered star formation (TSF) occurs when supersonic flows generated by distant supernova blast waves or stellar winds (wind blown bubbles) sweep over a stable cloud.  In realistic environments, this is likely to occur when such a flow impinges the heterogeneous regions within molecular clouds (\citet{robe69}, \citet{hill97}, \citet{koth01},  \citet{bonn06}, \citet{leao09}). While it is unclear if TSF accounts for a large fraction of the star formation rate within the galaxy, the concept has played an important role in discussions of the formation of our own solar system because it offers a natural way of injecting short lived radioactive isotopes (SLRI's) like $^{26}Al$ and $^{60}Fe$ into material which will then form planetary bodies.

In light of SLRI observations, a series of studies dating back to the 1970s (\citet{came77}, \citet{reyn79}, \citet{clay93}) have attempted to quantify the ability of a blast wave or stellar wind to both trigger collapse in a stable cloud and inject processed material. Because of the complex nature of the resultant flows, these studies have relied strongly on numerical simulations (\citet{boss95}, \citet{fost96}, \citet{vanh98}, \citet{vanh02}). In a more recent series of papers by Boss and collaborators (\citet{boss08}, \citet{boss10}, \citet{boss13}) the shock conditions needed for successful triggering and mixing were mapped out. In general, the higher the Mach number of the shock, the more difficult it is to trigger collapse. Faster shocks can shred and disperse the clump material before it has time to collapse. However faster shocks also allow better mixing by enhancing Rayleigh-Taylor instability growth rates. \citet{boss10} have shown that for a stable cloud of $1 M_{\odot}$ and radius of $0.058$ pc, the incoming shock needs to be slower than $80$ km/s to trigger collapse.  The shocks also need to be at least $30$ km/s to yield $10\%$ of blast material (by mass) mixing into the cloud.  Thus there is a relatively narrow window, in terms of shock Mach number, where both triggering and mixing can be achieved. 

\citet{boss102} and \citet{boss13} further pointed out that in order to explain the abundance of $^{26}Al$ in the Solar System using  triggering, the supernova shock needs to satisfy additional width requirements besides the shock speed condition. Finally, \citet{grit12} pointed out the importance of cooling in such a triggering scenario, detailing the condition for  collapse of the cloud fragments  by thermal instability. We note also \citet{harp14} who studied the possibility of forming low-metallicity stars by supernova shock triggering with simulations and \citet{vaid13} who studied the collapse of magnetically sub-critical cloud cores. 

These studies have done much to reveal the details of TSF but they have been restricted to the early stages of the resulting flow pattern.  The full evolution leading to a collapsed object (a star) and its subsequent gravitational interaction with the surrounding gas has yet to be studied. Part of the difficulty has been the numerical challenge of  generating a sub-grid model for the collapsing region that adequately represent stars. This has left  many questions unanswered.  For instance, what is the mass accretion rate of such a star formed by triggering? What is the accretion history of such a star? Does a trigger-formed star also have a disk when rotation in present in the cloud? If so, is the disk stable?  Some of these questions, such as disk stability, have been studied in other contexts: \citet{ouel07} explored disk ablation when the disk was swept over by a supernova blast wave and ejecta.  They found the disks to be long-lived and relatively stable in spite of the supernova blast impact. Their disks were not, however, formed by triggering but were considered to be pre-existing.  Determining the surviving disk mass and  the mixing between cloud and wind material is important for understanding the role of TSF  in Solar System formation and/or in supplying  SLRI abundances.

We note that the issue of triggering is of more general interest than discussions of SLRIs.  For example the in the HII regions associated with the Carnia nebulae a number of elongated pillars are seen with jets emerging from the head of the pillar (HH901 and HH 902 \cite{smithea2010}).  The presence of the jet is an clear indication of the presence of a newly formed star at the head of pillar.  If the pillars are formed via a combination of photo-ablation and winds from the massive star then one would expect shock triggering to occur within any marginally stable clumps in the pillar material once the shock reached the clump position.   Thus the dynamics of star formation within HII region pillars represents another of many reasons why TSF needs to be explored in its full evolutionary detail.

In this paper, we use the parallel AMR code AstroBEAR2.0 (\citet{cunn09}, \citet{carr13}) to study the shock-induced triggering of a stable Bonnor-Ebert cloud following, for the first time, the long-term evolution of the system after a star, numerically represented by a sink particle, has been formed.

To explore the post-triggering physics of TSF, we present simulations in three different regimes: I. triggering a non-rotating cloud; II. triggering a cloud with an initial angular momentum parallel to the shock normal; III triggering a cloud with an initial angular momentum perpendicular to the shock normal. These simulations allow us to answer four questions: 1. What is the nature of the flow pattern after a star has formed in TSF? 2. How do disks form in TSF environments?  3. what is the subsequent disk evolution in the presence of the post-shock flow?  4. How do accretion and mixing properties change with initial conditions in TSF? In particular we explore the evolution and the disruption of the protostellar envelope by the post-shock flow. For the rotating cases, we are interested in how the initial angular momentum  can lead to formation accretion disk surrounding the newly formed star.  Finally, we study the interaction of the disk and the post-shock flow and its affect on circumstellar disk survival.

The structure of this first report of our ongoing campaign of simulations is as follows.  In Section 2 and 3 we describe the numerical method and model. In Section 4 we report our results.  Section 5 provides analysis of the results and we conclude in Section 6.

\section{Hydrodynamic Equations with Self Gravity}
For our  simulations, we use  the AstroBEAR 2.0 code with a 3D computational grid. AstroBEAR is a parallel AMR magneto-hydrodynamics code with multi-physics capabilities that include self-gravity, micro-physics, heat conduction, resistivity, non-ideal EOS, and radiation transfer. Details on AstroBEAR may be found in \citet{cunn09}, \citet{carr13} and at https://clover.pas.rochester.edu/trac/astrobear. While the code can treat multiple atomic, ionic and molecular species, in this work we work under the single fluid assumption with tracers to track different types of material: cloud, initial ambient and post-shock. We assume uniform atomic mass $\mu_A = 2.3m_H$, where $m_H$ is the atomic mass unit, and employ an approximate isothermal assumption: $\gamma = 1.0001$. 

The Euler equations for density, momentum density and energy density that we solve numerically  are as follows:
\begin{equation}
\frac{\partial \rho}{\partial t}+\nabla \cdot (\rho \bf{v}) = 0,
\end{equation}
\begin{equation}
\frac{\partial (\rho \textbf{v})}{\partial t}+\nabla \cdot (\rho \textbf{vv} + p) = -\rho \nabla \phi,
\end{equation}
\begin{equation}
\frac{\partial E}{\partial t}+\nabla \cdot [\textbf{v}(E+p)] = -\rho \textbf{v} \cdot \nabla \phi,
\end{equation}
where $\rho$, $\textbf{v}$ and $p$ are the density, velocity and pressure, $E$ denotes the total energy density given by
\begin{equation}
E = \epsilon+p\frac{\textbf{v}\cdot \textbf{v}}{2},
\end{equation}
where the internal energy $\epsilon$ is given by 
\begin{equation}
\epsilon = \frac{p}{\gamma -1}
\end{equation}
where we take isothermal approximation: $\gamma = 1.0001$. $\phi$ is the gravitational potential given by the the Poisson equation: 
\begin{equation}
\nabla^2 \phi = 4 \pi G \rho.
\end{equation}
At each simulation step, we first predict the distribution of the gravitational potential by solving Eq.(6) using linear solver package HYPRE (\citet{bake12}). We then solve the fluid equations with the external gravitational force field as shown in Eq.(2) and Eq.(3) with the MUSCL (Monotone Upstream-centered Schemes for Conservation Laws) primitive method using TVD (Total Variation Diminishing) preserving Runge-Kutta temporal interpolation. For details of the implementation and tests of self-gravity in AstroBEAR see \citet{kami14}. AstroBEAR shows excellent scaling up to $10^4$ processors \citet{carr13} and the simulations presented in this paper were carried out on the Kraken cluster of NICS on 1200 cores.

The formation of the star is treated numerically by the introduction of a sink particle. When the local fluid quantities satisfy  a set of criteria ensuring collapse within a simulation cell, a particle of appropriate mass is created and subsequently treated as a point gravity source. This zero-dimensional particle, once formed, moves through the grid via gravitational interactions with its surrounding gas and other particles. Several numerical schemes have been employed to calculate the conditions for sink particle formation as well as the rate at which the particle accretes gas (\citet{krum04}, \citet{fede10}, \citet{gong13}). In this paper, we form particles based on criteria proposed by \citet{fede10}, and then calculate the accretion of the particle based on the accretion scheme described in \citet{krum04}. The reasons for choosing the Krumholz accretion scheme over that of Federrath is related to our use of the isothermal approximation and will be discussed further  in Section 4.

\section{Initial Simulation Setup}
We begin with an initial marginally stable Bonnor-Ebert sphere as the triggering target for our shock. The initial cloud setup is similar to \citet{boss10}, i.e a cloud with $M_c = 1 M_{\odot}$, a radius of $R_c = 0.058 pc$, a central density of $\rho_c=6.3\times10^{-19} {\rm g/cc}$ and edge density of $3.6\times 10^{-20} {\rm g/cc}$. The cloud has a uniform interior temperature of $10K$. The ambient medium is initialized to satisfy  pressure balance at the cloud boundary when the cloud is stationary, with density $\rho_a = 3.6\times 10^{-22} {\rm g/cc}$ and temperature of $1000 K$. 

We express time scales in terms of the ``cloud crushing time" $t_{cc}$ which is defined as the time for the transmitted shock to pass across the cloud, i.e. $t_{cc} = \sqrt{\chi} R_c/V_s$ where $V_s$ is the incident shock velocity and $\chi \approx 1700$ is the ratio of peak cloud density to ambient density.  For our conditions $t_{cc} \approx 276$ kyrs. We have performed simulations to check the stability of the cloud and find that the cloud oscillates with a time scale of about $10t_{cc}$.  This is longer than the time span of our simulation. The free-fall time $t_{ff}$ can be used to gauge the time scale of gravitational collapse. Our initial cloud has $t_{ff} \approx 84$ kyrs. Note that although we find that triggering can form a star as early as  $t_{cc}<t< 2t_{cc}$, our interest in the post-triggering interaction leads us to simulate the fluid evolution through $4t_{cc}$, which is approximately equivalent to $1$ million years. To make a comparison between slow and fast shock cases, we initialize the incoming shock at two different Mach numbers: either $M = 1.5$ or $M = 3.16$, where $M$ is the ratio between the shock speed and ambient sound speed: $M = v_s /c_s$. Given the shock speed $v_s = 3 km/s$, we can estimate the incoming mass flux as $F_s = 4\pi \rho_a v_s \approx 1.4 \times 10^{-13} g/cm^2s $.

We use $K = \Omega t_{ff}$ to characterize the importance of rotational energy in our simulations where $\Omega$ is the angular velocity. We assume $K = 0.1$ for all the rotational cases presented in this paper (\citet{bane04}). Characterizing the influence of different $K>0$ values is an important separate topic that we leave for future work.
Here we simply focus  on studying the difference between the rotating ($K = 0.1$) and non-rotating cases ($K = 0$) and  different orientations of the initial rotation axis. Adding initial rotation can change the initial equilibrium of the cloud, but we have performed simulations to verify that for $K < 0.5$, the cloud remains stationary within the simulation time span. The parameters of the initial setup are summarized in Table.1. 

We continue to inject a "post-shock wind" of the same form as that used in \citet{boss10} (and of the same density and temperature as the initial ambient gas) until the end of the simulations  (i.e. long after the initial shock has passed by the cloud). We assess how strongly this wind ablates the bound cloud material,  including that material which forms a disk in the rotational cases. The density of this post-shock wind is approximately $100$ times lighter compared to the shock front, giving a mass flux of $F_w \approx 1.4 \times 10^{-15} g/cm^2s $. 
Although  continuation of this post-shock wind for the full duration of the simulation is unphysical because  it implies a total mass loss of $198M_{\odot}$ ejected from a source $1$pc away,  it will tell us that any disk which  survives this extended wind will  also survive any shorter lived wind with the same mass flux .\\

\begin{table}
\caption{Simulation Setups}
\label{tab01}
\begin{tabular}{|c|c|c|c|}
\tableline
Code & Shock Mach & Cloud Rotation (relative to shock normal) & K   \\
    N  &      $1.5$    &        None        &0.0 \\
    N' &      $3.16$   &        None        &0.0 \\
   R1 &      $1.5$     &      Parallel       &0.1 \\
   R2 &      $1.5$     &  Perpendicular  &0.1 \\
\tableline
\end{tabular}\\
\end{table}

We implement mesh refinement to focus on the region centered on the sink particle. The simulation box has a base resolution of $320 \times 192 \times 192$, which is equivalent to $64$ cells per cloud radius. We add $3$ levels of refinement around the  region of the cloud (or sink particle) yielding an effective resolution of $64 \times 2^3 = 512$ cells per cloud radius. We employ outflow boundary condition at all the boundaries of the simulation box. \\

\section{Simulation Results}
In general, we can divide the triggering event into three phases: I. the incoming shock impinges on the cloud compressing it into a dense core until the local Jeans' stability criterion is violated.  The subsequent infall generates a star (represented by a sink particle in our simulations)  marking the end of this phase. II. Ablated cloud material that is not gravitationally bound is accelerated and ejected downstream.  The  still gravitationally bound gas is also  exposed to the post-shock wind.  III. The star and its bound material continue to evolve while interacting with the post-shock wind until the end of the simulation. 

Fig.1 demonstrates these stages.  In the figure we show the column density (density integrated along the axis pointing out of the plane) evolution of case R1 (see Table 1)  immediately after the star is formed, at about $1.1t_{cc}$ ($0.3$ million years) in the top panel; immediately after the star has entered the post-shock region ($0.5$ million years) in the middle panel; and after the star and its surrounding disk become embedded completely in the post-shock wind in the bottom panel. In Fig.1(a),  a star (represented by a red sphere) embedded in the cloud is visible as the collapse proceeds. In Fig.1(b), the star, as well as the bound cloud material has been left behind as the unbound remnant cloud material is driven downstream (to the right).  The star and the gas bound by its gravitational potential remain exposed in the post-shock wind.  At this point, the  initial angular momentum of the cloud (oriented along the shock normal) leads to the creation of a disk. In Fig.1(c), we capture the flow pattern at  time $\sim 0.85$ million years. Here,  although the disk has  experienced a ram-pressure driven ablation from the post-shock flow for more than $0.3$ million years, its shape and size remain relatively unchanged.  As noted in Section 3, in reality the post-shock flow will last less than $1$ million years, so our results conservatively indicate that disks should survive the post-shock environment of a typical triggering event. This survival is discussed in more detail in Section 5.4. \\

\begin{figure}
\includegraphics[scale=0.3]{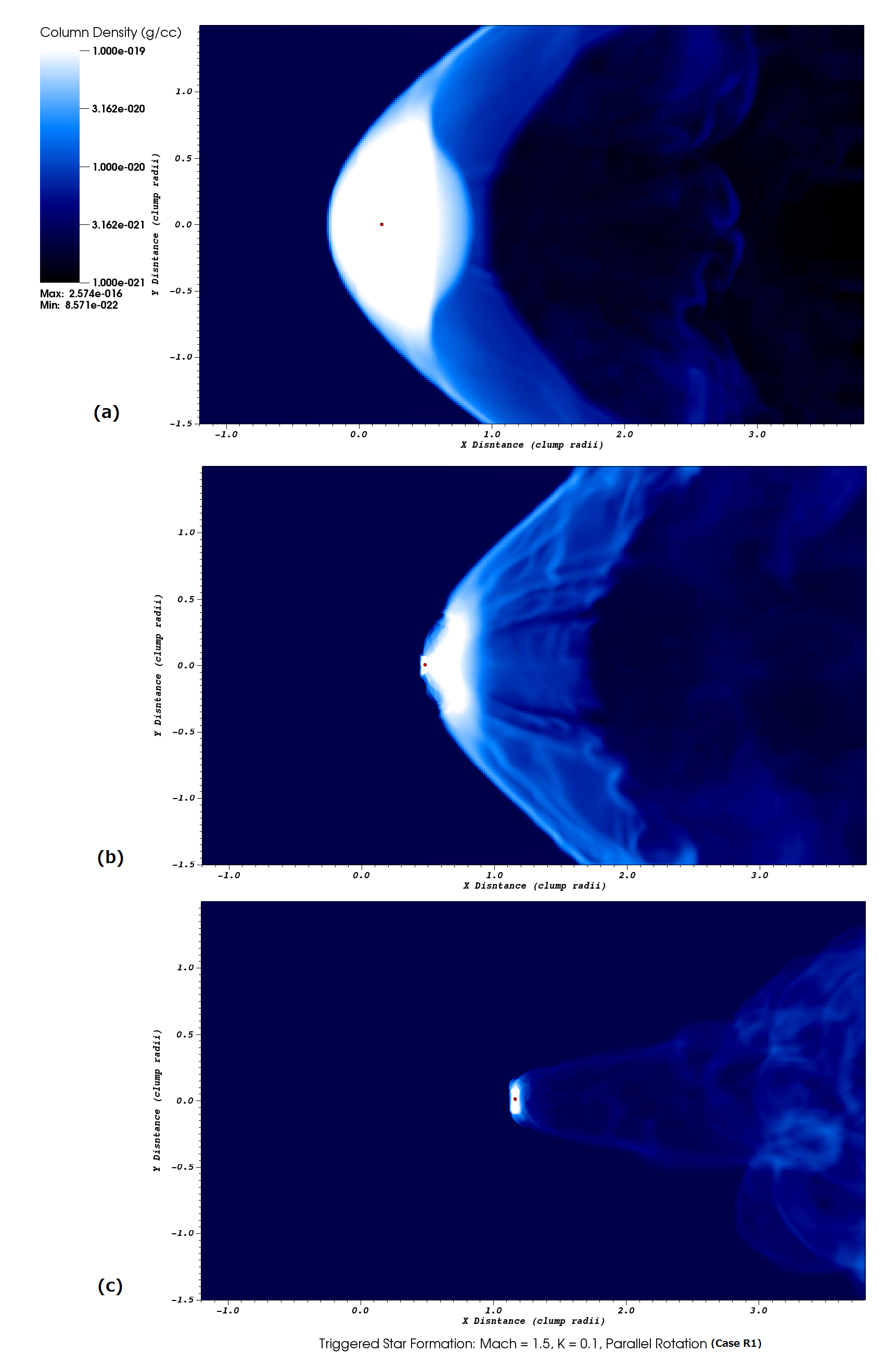}
\caption{Column Density Evolution for Case R1: (a) 0.3 million yrs; (b) 0.5 million yrs; (c) 0.85 million yrs}
\label{fig01}
\end{figure}

To compare the different cases listed in Table 1, in Fig 2 we plot the column density of each case at a fixed time  - $0.6$ million years. This corresponds to just after the star has entered the post-shock wind, and the disk, if it forms, is present. For case N, the  bound  cloud material surrounding the newly formed star  is quickly shredded away by the post-shock flow, leaving the star isolated in the wind. Given the low density of the resulting circumstellar material,  its accretion rate is low and the bulk of mixing be determined before the end of phase II. 

For case N', the incoming shock is approximately twice as fast as that in case N. We observe that star formation can still be triggered, confirming that  $Mach = 3.16$ falls in the ``triggering window" (less than Mach $20$) described in \citet{boss10}. The time scale for the triggering $t_t$, defined as the time scale between the beginning of the shock compression until the formation of the star,  is half of that of case N. \\

\begin{figure}
\includegraphics[scale=0.15]{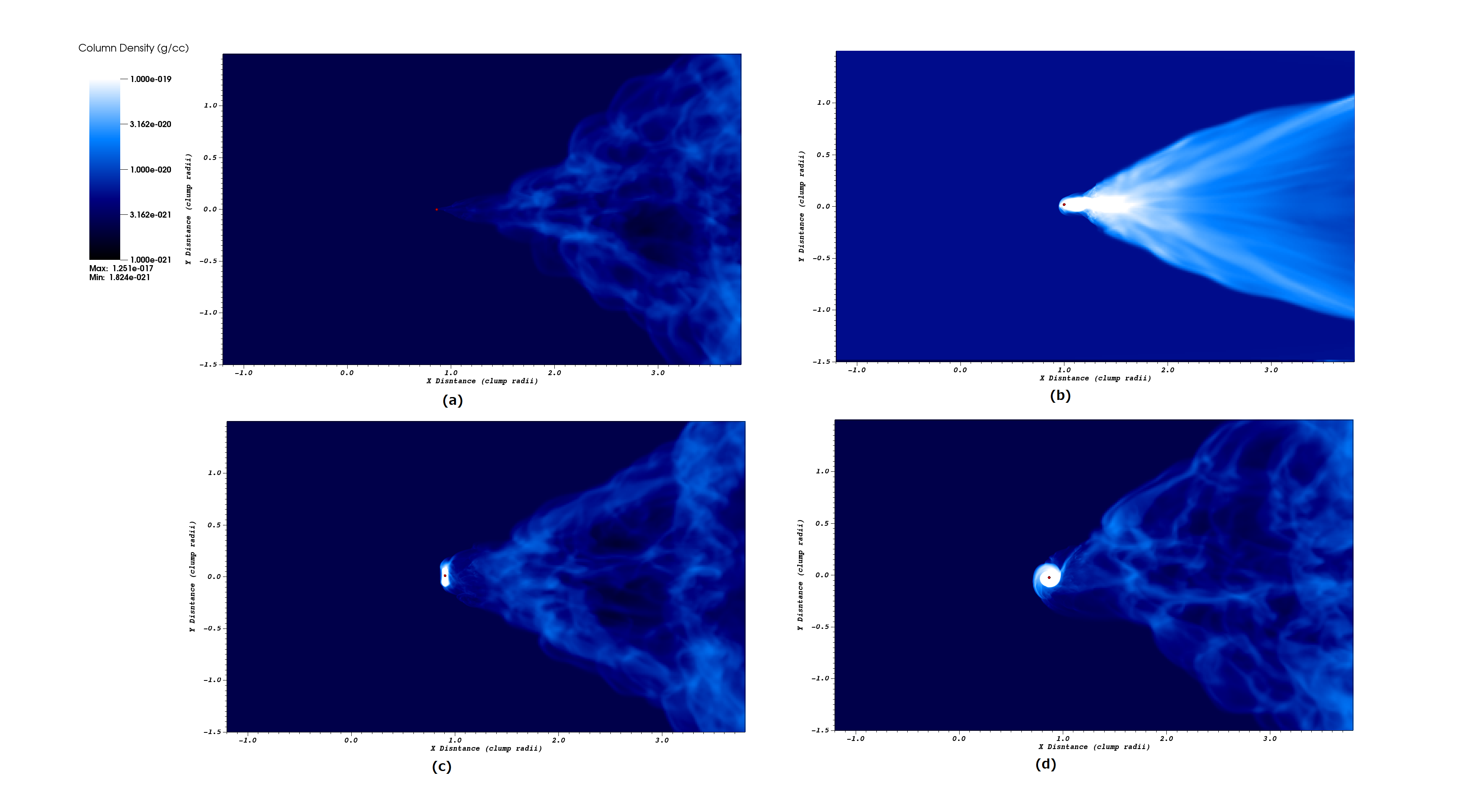}
\caption{Post-triggering evolution at $0.6$ million years: (a) Case N; (b) Case N'; (c) Case R1; (d) Case R2. }
\label{fig02}
\end{figure}

For cases R1 and R2, the bound material forms a disk of radius  $\sim 1000 $AU at the end of phase II. This disk radius is consistent with the estimation of disk formation radius $r_d \approx \Omega^2 R_c^4/2 G M_s$, where $M_s$ is the mass of the central star (about $1 M_{\odot}$). This expression for $r_d$ is determined by the radius at which material  in-falling 
while  conserving angular momentum reaches a Keplerian rotation speed. Note also that the disk temperature deviates from the initial cloud temperature ($10$K) because $\gamma$ is set to $1.0001$ instead of exactly $1$. For Federrath type accretion algorithm, this temperature increase can introduce heated numerical accretion zone (a zone of fixed number of cells that are kept at just below the threshold density) around the star. Once the star drifts into the post-shock wind, this heated zone can expand and disrupt the circumstellar profile. This is the reason why we preferred to choose Krumholz accretion algorithm which does not rely on creating such an accretion zone. We have verified through simulations that when $\gamma-1$ is approaching zero, the triggered star formation results obtained from Federrath and Krumholz type accretion algorithms converge.

The disk formation is a natural consequence of the initial rotation, as in both cases the planar shock does not significantly alter the angular momentum distribution of the cloud as long as the shock remains stable. In the the N cases, little post-shock circumstellar material remains compared  to the R cases since the material can more easily collapse to the core for the former cases. However,  the total post-shock stellar plus bound circumstellar material is lower for the R cases than the N cases since the presence of angular momentum makes material less tightly bound initially.

\begin{figure}
\includegraphics[scale=0.24]{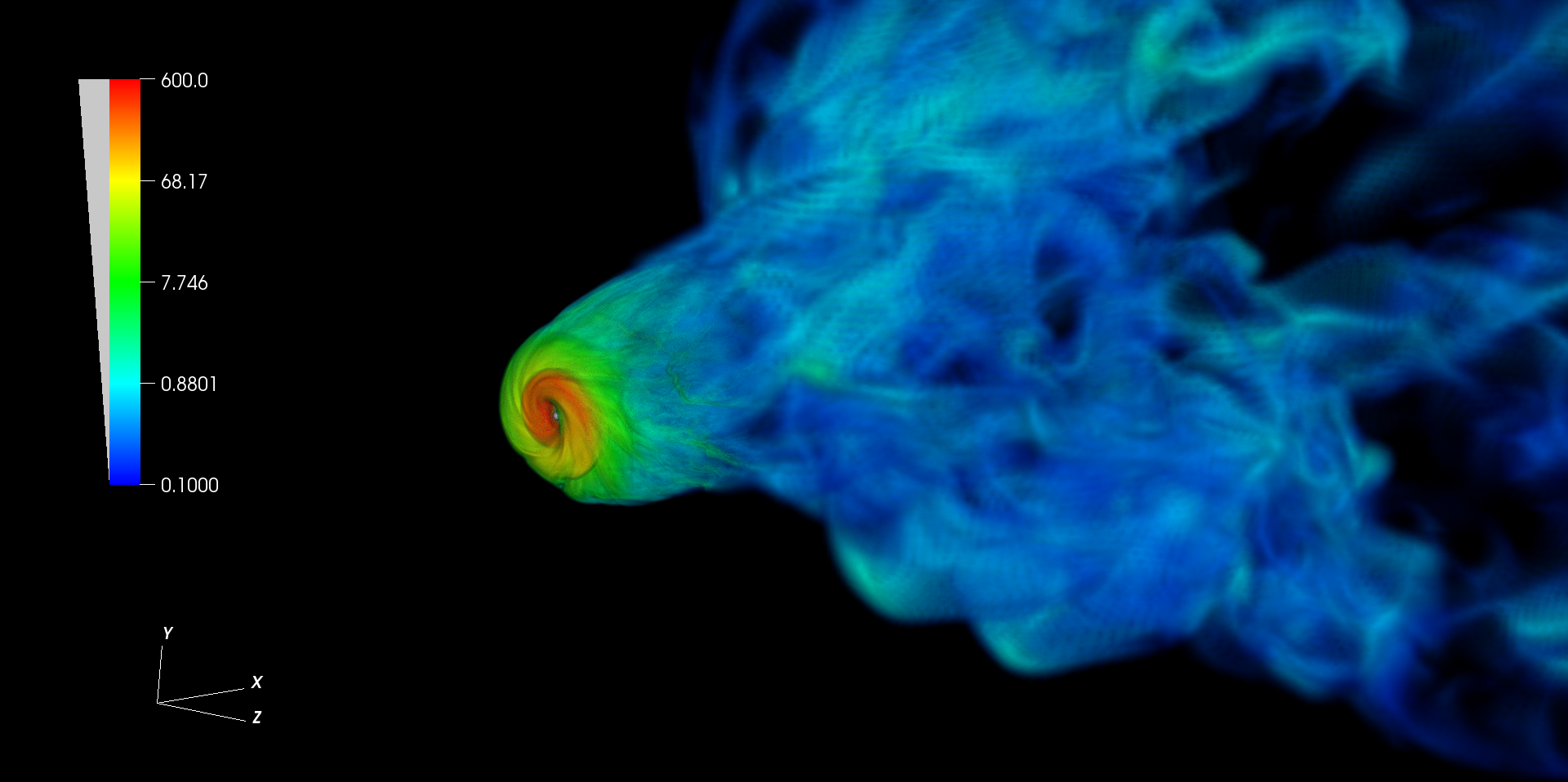}
\caption{Case R1: 3D volume rendering of the disk formed by triggering at $0.6$ million years. }
\label{fig03}
\end{figure}

We also expect less mixing in the N cases compared to the R cases given the same shock Mach number because an extended disk acts to trap some of the incoming material. But  because R1 and R2 have different orientations of the disk relative to the incoming wind, we expect the mixing of material into the disks in these two cases to also be different.  In case R1, the  disk presents the maximum cross section for ablation ($\pi r_d^2$)  while in case R2 , the wind hits the disk edge on, yielding a much smaller cross section $\propto h$ the vertical scale height.  Case R2 exhibits an ellipsoidal disk geometry  just after its formation due to the disk-wind interaction.   In short, comparing the R and N cases,  we can qualitatively understand the differences in both accretion rates and mixing ratios.

In Fig.3, we plot a 3D volume rendering of case R1, at time $0.6$ million years. This corresponds to the time period after the disk has been completely engulfed in the post-shock wind. The pseudo-color shows the density percentage as normalized by the initial average cloud density - initial average cloud density is set as 100. Fig 3 shows that the compressed cloud material (red region in Fig.3) mostly ends up accreted onto the star (marked in Fig.3 as the white sphere) or in the accretion disk. The figure shows the spiral pattern that forms downstream as  disk material is ablated by the post-shock flow.\\

\section{Quantitative Discussion}

In this section we briefly discuss the implications of our simulations, in terms of the physics of triggering and subsequent star/disk evolution, given the cases we have studied.  We saved a more complete exploration of parameter space and its astrophysical implications for future work. 

\begin{figure}
\includegraphics[scale=0.24]{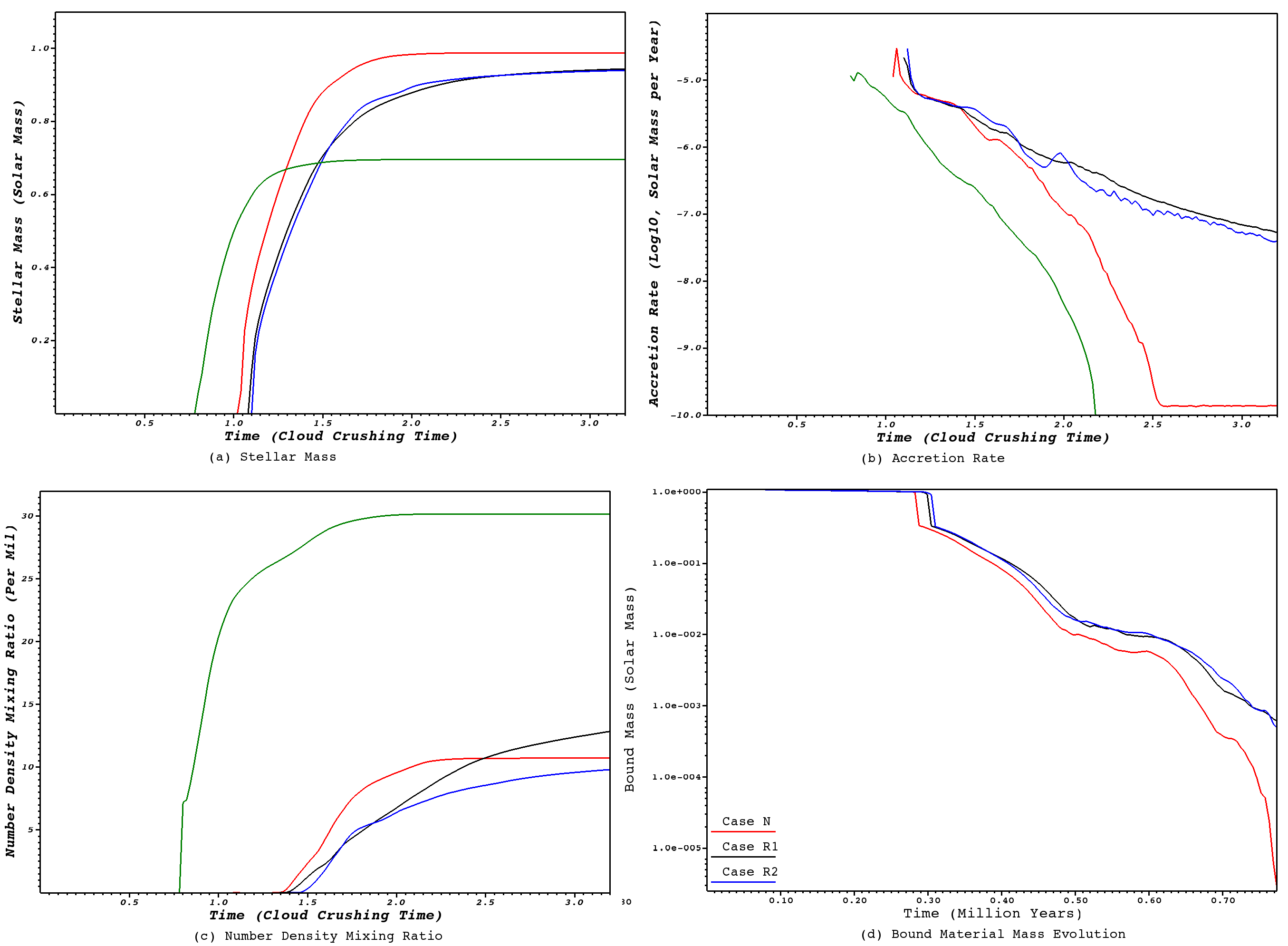}
\caption{Time Evolution of  Stellar Mass, Accretion Rate, Wind Material Mixing Ratio and Bound Mass}
\label{fig04}
\end{figure}

\subsection{Triggering time}

 In Fig.4 (a), we plot the evolution of the stellar mass (represented by sink particle mass) formed by the triggering event for the four cases. Note first that in all four simulations the star forms at around $0.8$ to $1.2 t_{cc}$, which corresponds to about $0.2$ to $0.3$ million years for the Mach $M=1.5$ cases, and about $0.12$ million years for the  $M=3.16$ case. Case N' has an absolute formation time of about half of that of Case N, due to its fast compression. For the transmitted shock, the density compression ratio $\eta$ is related to the transmitted shock Mach number $M$ via $\eta \propto M^2$. This is because the force exerted on the cloud is proportional to the ram pressure of the incoming wind $\rho_w v_s^2$, where $\rho_w$ is the wind density defined in Section 3 and $v_s$ is the shock velocity: $v_s = M c_s$. If we assume that the compressed cloud material behind the transmitted shock undergoes free-fall collapse, we can estimate the collapse time scale as $t_{ff} \propto 1/\sqrt{\eta}$. This yields a scaling  for the triggering time described in the last section as $t_t \propto 1/M$.  If the triggering time is inversely proportional to force on the cloud, then  as we increase the Mach number by a factor of $2$ as occurs in the set up of  Case N vs. Case N', we expect the triggering time to be approximately halved. This is consistent with Fig.4(a).  The rotating cases R1 and R2 have slightly later triggering times compared to the non-rotating cases, because of the additional support against collapse provided by the added rotation.  When $K$ is small, the inward acceleration is reduced by $\Omega R^2$, where $R$ is the orbital radius of the considered gas parcel. The in-fall time is then calculated from:
\begin{equation}
\frac{1}{2}(GM/R^2 - \Omega^2 R) t_{in}= R
\end{equation}
from relations $GM/R^2 = 2R/t_{ff}^2$ and $\Omega^2 R = R K^2/t_{ff}^2$, we obtain the in-fall time is increased as $t_{in} = \sqrt{1+K^2} t_{ff}$ when initial rotation is added. The delayed triggering time can then be seen as the effect of the  $K^2$ term. \\
 
 \subsection{Asymptotic Stellar Mass}
 
Another significant feature shown in all four cases is the asymptotic stellar mass found in the simulations. We find  $M_* \sim 1M_{\odot}$ for the Mach $1.5$ cases, and $0.6M_{\odot}$ for the Mach $3.16$ case. The lower asymptotic mass of case N' can be explained by the fact that once a sink particle is formed, its accretion rate is  determined by the Bondi accretion rate implemented through the \citet{krum04} accretion algorithm. Thus, the stellar mass at the end of phase I, and consequently the asymptotic stellar mass, is predominantly determined by how much time the particle has to accrete cloud material before it enters the post-shock wind region.   This time scale is determined by how fast the incoming shock can accelerate the cloud material. Using the analysis of \citet{jone96} we have the ``cloud displacement" time  $t_{dis} = \sqrt{R_c/a_c}$, where $R_c$ and $a_c$ are the cloud radius and acceleration, respectively. Since $a_c$ is proportional to the ram pressure from the shock exerted on the cloud, $a_c \propto M^2$. This yields a time scale $t_{e} \propto \sqrt{R_c}/M$ for the star and its bound material to become exposed to the post-shock.  Thus case N' has about half the time to accrete cloud material as compared to cases N, R1 and R2. Fig.4(a) agrees with the above analysis. For the Mach $1.5$ cases, the final stellar mass approaches $M_* \sim 0.98 M_{\odot}$ for the non-rotating case, and $M_* \sim 0.94 M_{\odot}$ for the two rotating cases. This indicates that for all the cases studied, most of the initial cloud material ends up in the star before the end of phase II, which is consistent with the discussion in Section 4. 

The reduced stellar mass for the rotating cases is reasonable as some of the material ends up in a disk as opposed to directly accreting onto the star. At the end of stage I ($0.45$ million years for the R cases), the gravitationally bound gas enters the post-shock region, and the disk is visible in  the simulations. This disk has an initial mass of approximately $0.1M_{\odot}$, which is in agreement with the initial $K$ and its radius as discussed in the previous section. The disk mass gradually depletes because of the accretion onto the star as shown in Fig.4(d), and the stellar mass continues to increase during stage II for the R cases. At the end of the simulation, the disk mass drops to less than $10^{-3}M{\odot}$. We will discuss the wind ablation and the asymptotic disk mass in more detail in Section 5.4. \\

 \subsection{Accretion Rates}
In Fig.4 (b), we present the stellar accretion rates in our models. The accretion rate is calculated as the time derivative of the stellar mass. The most conspicuous feature is the difference between the non-rotating and rotating cases. While case N reaches its final accretion rate at approximately $0.7$ million years (set by Bondi-Hoyle accretion in the post-shock flow), cases R1 and R2 continue to accrete mass at a  higher rate because the mass was unable to fall in earlier and is in the disks.  The higher accretion rate at these times for the R cases can be thought of as  ``delayed"  infall: in the R cases, some of the cloud material ends up in the disk instead of being immediately accreted by the star due to the additional support provided by the rotation. This material can still be accreted through the disk later in stage II (i.e. accretion is delayed). The total mass that becomes the star would be
 is overall less for the R cases.

The disk formation and subsequent accretion  aids in mixing more material from the shock (and post-shock) gas into the star compared to previous studies without such disks as the disk provides greater cross section for interaction with the incoming wind. The accretion efficiency of wind material during stage II is set by the cross section of the total bound gas embedded in the wind (star+gas).  This cross-section is $\pi r_d^2$ for the R cases. For the N cases it is determined by the  Bondi radius: $\pi r_B^2$  where $r_B = 2GM_*/(c_s^2 + v_w^2)$. Given the parameters $M_* \approx M_{\odot}$, $T = 10K$ and $v_w = 3 km/s$, we find  that $r_d^2 \gg r_B^2$.

We define the mixing ratio as the ratio of $\kappa = n_w/(n_c+n_w)$, where $n_w$ and $n_c$ are the number densities of the post-shock gas and cloud gas that end up accreted onto the star, respectively. In Fig.4 (c), we see that the parallel rotation case has the highest mixing ratio amongst the three Mach $1.5$ cases.  As discussed earlier, this is likely due to its large cross section of interaction with the post-shock flow. Case R2 has a lower mixing ratio compared to N at the end of the simulation but the R2 rate is still growing while the N rate has reached its maximum value. Note that the $M=3$ case shows much more mixing than  the lower Mach number simulations. This is likely the result of increased shock speed on the internal flow within the cloud and is consistent with \citet{boss08}, where the effect of shock Mach number on mixing ratio was more thoroughly explored.\\

\subsection{Circumstellar Bound Mass and Disk Survival}

Finally in Fig.4 (d), we present the mass evolution of the cicumstellar gravitationally bound gas where we label any gas parcel with total energy $E = E_k+E_{th}+E_{gas-gas}+E_{gas-particle} < 0$ as bound ($E_k$ is the kinetic energy, $E_{th}$ is the thermal energy, $E_{gas-gas}$ and $E_{gas-particle}$ are the gravitational binding energy from self gravity and the star's point gravity.) The initial kink in the three curves at around $0.3$ million years coincides with the onset of triggering. From $0.3$  to $0.5$ million years, the shapes of the curves remain similar. This is in phase I where the star has not yet emerged from the cloud, and most of the mass loss results from the accretion onto the star. 

Since case N does not form a disk, the circumstellar bound material is quickly shredded away by the incoming wind once exposed to the post-shock flow. At $0.8$ million years, its bound mass drops to about $100$ times less than that of the two rotating cases. There is no resolvable material left surrounding the formed star. For cases R1 and R2, the bound mass drops at a much slower rate because of the disk. From Fig.4 (d), we observe that if the wind is turned off prior to $0.7$ million years, the surviving disk will have a mass greater than $10^{-3}M_{\odot}$, giving the mass of the whole system $1.001 M_{\odot}$, close to the Solar System. Therefore we conclude that it is possible to obtain at least a $1.0014M_{\odot}$ star plus protoplanetary disk system from such a triggering mechanism given our physically reasonable choice of  initial conditions.

To connect our disk survivability results with previous work,  we follow \citet{chev00} and estimate the erosion radius $r_e(t)$ of the disk from $\rho_d(t) \sqrt{2GM_*/r_e(t)} = \rho_w v_w$, where $\rho_w$ and $v_w$ are the density and velocity of the post-shock wind and $\rho_d(t)$ is the density of the disk. Material at radii $r> r_e(t)$ cannot survive in the disk assuming that the wind momentum is fully transferred to the disk.   Any disk   surviving at a given time must have   $r_d (t)< r_e(t)$.  For our simulations we have $\rho_d \approx 10^{-18} g/cc$, $\rho_w = 3.6 \times 10^{-22} g/cc$, $M_* \approx M_{\odot}$ and $v_w = 3 km/s$ at the end of our simulations thus we can verify that $r_d \ll r_e$.
Although this is a necessary property that a surviving disc  must have at the end of the simulation,  the condition evaluated at the {\it initial time}  of disk formation is not  sufficient  to assess its  long term survivability because it does not account for the accumulated influence of the wind.  Even a low density wind impinging over long enough times could in principle ablate the disk.   However  our disk survival  is also in agreement with the study by \citet{ouel07}, who found that pre-existing disks can survive ablation from the full exposure to supernova driven shock. Such survival can only result if   the drag of the disk on the wind is  inefficient. Indeed  \citet{ouel07} find that a high pressure region and reverse shock formers upstream of the disk surface and deflects the flow around the disk leaving it intact. The result is that the  wind-disk interaction is ineffective at disk ablation.\\
 
\section{Conclusion}

Using AMR numerical simulations, we have followed the interaction between shocks of different Mach numbers and self-gravitating clouds, with and without initial rotation. In each case we followed the evolution of  the interaction to study collapse of the cloud,  formation of a star, and  post-shock  evolution as the wind continues to interact with the collapsed cloud.   Our studies have carried out the shock-cloud interaction  to longer  times than  have been previously studied. Our focus has been on the extent to which the variation in Mach number and the presence of rotation (at 10\% the escape speed) affects star formation, the post-collapse circumstellar  bound  mass,  and the mixing of blast wave material with the cloud. In all three cases that we studied, the interaction proceeds in three phases. First the shock compresses the cloud enough to form a star at the core. Then some cloud material gets ablated and unbound from the star. Finally, some  material remains bound to the star and continues to evolve as it is exposed to the post-shock flow. The star formation from the shock induced collapse is robust in all cases whether rotating or not. The mass of the star formed in the initial collapse phase is also comparable in the rotating and non-rotating cases but slightly larger in the non-rotating case since the rotation makes the total mass less bound than for the non-rotating case.  However the shock Mach number affects the asymptotic stellar mass even more than the rotation: the higher the Mach number, the less the stellar mass at the end of the simulation. 

For the case of rotating clouds, bound circumstellar disks form around the newly formed stars.
Even though the disks are  exposed to a continuous stellar wind for throughout the long duration of our simulations,  the disk  survives this long duration of wind erosion. Because the net  momentum from the wind impinging on the disk  is substantial, the survival of the disk implies that the drag on the wind by the disk is small, leading to inefficient conversion of the full wind momentum to disk ablation flow. Overall,  the asymptotic disk mass of  around $10^{-3} M_{\odot}$ given our 1 $M_\odot$  initial cloud, is achieved when the  wind duration at  $~0.7$ million years.

For the question of mixing, we find that the dominant influence on the  mixing ratio of blast wave to  bound cloud material  is the Mach number of the initial shock. The higher the Mach number, the higher the mixing ratio.  The mixing ratio is relatively insensitive to the rotation. We note however that rotation can lead to disk formation which subsequently increases the cross section of the bound mass around the star and that can favor extra trapping of incoming wind material (when comparisons are made at a given Mach number with and without rotation).

Based on previous studies of Boss and collaborators that explored the relation between SLRI mixing and incident shock mach numbers, the simulations we present here (with M = 1.5 or 3) are not high enough to yield sufficient injection of material to account for observed SLRI abundences.  Given the earlier work we would need  Mach numbers in the range of 10 to 20 and we leave a fuller exploration of parameter space to a future work.  The simulation results presented here however do provide a general understanding to the long term evolutionary mechanisms of TSF including the effects rotation. \\

\acknowledgments{}
We thank John Carroll-Nellenback for his important contributions to discussions in this paper.  We also thank Baowei Liu, Erica Kaminski, Eddie Hansen and Zhou Chou for their help and insight. Support for this work was in part provided by NASA through awards issued by JPL/Caltech through Spitzer program 20269, the Department of Energy through grant number DE-SC-0001063, the National Science Foundation through grants AST-0807363 as well as the Space Telescope Science Institute through grants HST-AR-11250 and HST-AR-11251.  We also thank the University of Rochester Laboratory for Laser Energetics and funds received through the DOE Cooperative Agreement No. DE-FC03-02NA00057.  This research was also supported in part by the Center for Research Computing at the University of Rochester as well as the National Science Foundation through TeraGrid resources provided by the National Center for Supercomputing Applications. \\

\clearpage

\begin{thebibliography}{}

\bibitem[Baker et al (2012)] {bake12}
Baker, A. H., Falgout R. D., Kolev, Tz. V., Yang U. M., 2012, in High Performance Scientific Computing: Algorithms and Applications, eds. Springer, Scaling hypre's Multigrid Solvers to 100,000 Cores, LLNL-JRNL-479591

\bibitem[Banerjee et al (2004)] {bane04}
Banerjee, R., Pudritz, R. E., Holmes, L., 2004, \mnras, 355, 248

\bibitem[Bonnell et al (2006)] {bonn06}
Bonnell, I. A., Dobbs, C. L., Robitaille, T. P., Pringle, J. E., 2006, \mnras, 365, 37

\bibitem[Boss (1995)] {boss95}
Boss., A. P., 1995, \apj, 439, 224

\bibitem[Boss et al (2008)] {boss08}
Boss, A. P., Ipatov, S. I., Keiser, S. A., Myhill, E. A., Vanhala, Harri A. T., 2008, \apj, 686, 119

\bibitem[Boss et al (2010)] {boss10}
Boss, A. P., Keiser, S. A., Ipatov, S. I.,Myhill, E. A., Vanhala, Harri A. T., 2010, \apj, 708, 1268

\bibitem[Boss et al (2010)] {boss102}
Boss, A. P., Keiser, S. A., 2010, \apj, 717, 1

\bibitem[Boss et al (2013)] {boss13}
Boss, A. P., Keiser, S. A., 2013, \apj, 770, 51

\bibitem[Cameron et al (1977)] {came77}
Cameron, A. G. W., Truran, J. W., 1977, Icarus, 30, 447

\bibitem[Carroll-Nellenback et al (2013)] {carr13}
Carroll-Nellenback, Jonathan J., Shroyer, Brandon  Frank, Adam, Ding, Chen, 2013, J. Comp. Phy., 236 461

\bibitem[Chevalier (2000)] {chev00}
Chevalier, R. A. 2000, \apj, 538, 151

\bibitem[Clayton et al (1993)] {clay93}
Clayton, Donald D., Hartmann, Dieter H., Leising, Mark D., 1993, \apj, 415, 25

\bibitem[Cunningham et al (2009)] {cunn09}
Cunningham, A.J., Frank, A., Varni{\`e}re, P., Mitran, S., Jones, T.W., 2009, \apjs, 182 519

\bibitem[Dhanoa et al (2014)] {harp14}
Dhanoa, H., Mackey, J., Yates, J., 2014, http://arxiv.org/abs/1402.1103

\bibitem[Federrath et al (2010)] {fede10}
Federrath, C., Banerjee, R., Clark, P. C., Klessen, R. S., 2010, \apj, 713, 269

\bibitem[Foster et al (1996)] {fost96}
Foster, P. N., Boss, A. P., 1996, \apj, 468, 784

\bibitem[Gritschneder et al (2012)] {grit12}
Gritschneder, M., Lin, D. N. C., Murray, S. D., Yin, Q. Z., Gong, M. N., 2012, \apj, 745, 22

\bibitem[Gong et al (2013)] {gong13}
Gong, Hao,  Ostriker, Eve C., 2013, \apjs, 204, 8

\bibitem[Hillenbrand (1997)] {hill97}
Hillenbrand, L.A., 1997, \aj, 113, 1733

\bibitem[Jones et al (1996)] {jone96}
Jones, T. W., Ryu, Dongsu, Tregillis, I. L., 1996, \apj, 473, 365

\bibitem[Kaminski et al (2014)] {kami14}
Kaminski, E., Frank, A., Carroll-Nellenback, Jonathan J., Myers, P., submitted to \apj, http://arxiv.org/abs/1401.5064

\bibitem[Kothes et al (2001)] {koth01}
Kothes, R., Uyaniker, B., Pineault, S., 2001, \apj, 560, 236

\bibitem[Krumholz et al (2004)] {krum04}
Krumholz, M. R., McKee, C. F., Klein, R. I., 2004, \apj, 611, 399

\bibitem[Leao et al (2009)] {leao09}
Leao, M. R. M., de Gouveia Dal Pino, E. M., Falceta-Goncalves, D., Melioli, C.; Geraissate, F. G., 2009, \mnras, 394, 157

\bibitem[Li et al (2014)] {li14}
Li, S., Frank, A.,  Blackman, E.G., in prep.

\bibitem[Mac Low et al (1994)] {macl94}
Mac Low, M., McKee, C.F., Klein, R.I., Stone, J.M., Norman, M.L., 1994, \apj, 433, 757

\bibitem[Ouellette et al (2007)] {ouel07}
Ouellette, N., Desch, S. J., Hester, J. J., 2007, \apj, 662, 1268

\bibitem[Reynolds et al (1979)] {reyn79}
Reynolds, R. T., Cassen, P. M., 1979, Geo. Rev. Lett., 6, 121

\bibitem[Roberts (1969)] {robe69}
Roberts., W. W., 1969, \apj, 158, 123

\bibitem[Smith et al.(2010)]{smithea2010} Smith, N., Bally, J., 
\& Walborn, N.~R.\ 2010, \mnras, 405, 1153 

\bibitem[Walch et al (2013)] {walc13}
Walch, S., Whitworth, A. P., Bisbas, T. G., Wünsch, R., Hubber, D. A., 2013, \mnras, 435, 917

\bibitem[Vanhala et al (1998)] {vanh98}
Vanhala, Harri A. T., Cameron, A. G. W., 1998, \apj, 508, 291

\bibitem[Vanhala et al (2002)] {vanh02}
Vanhala, Harri A. T., Boss, A. P., 2002 \apj 575, 1144

\bibitem[Vaidya et al (2013)] {vaid13}
Vaidya, B., Hartquist, T. W., Falle, S. A. E. G., 2013, \mnras, 433, 1258

\bibitem[van Loo et al (2007)] {vanl07}
van Loo, S., Falle, S. A. E. G., Hartquist, T. W., Moore, T. J. T., 2007, \aap, 471, 213

\end{thebibliography}
\end{document}